\documentclass[twocolumn,english,showpacs,prl]{revtex4}
\usepackage[T1]{fontenc}
\usepackage[latin9]{inputenc}
\usepackage{verbatim}
\usepackage{amsmath}
\usepackage{amssymb}
\usepackage{graphicx}
\usepackage{amsfonts}
\usepackage{dcolumn}
\usepackage{bm}
\usepackage{babel}

\makeatletter

\@ifundefined{textcolor}{}
{
 \definecolor{BLACK}{gray}{0}
 \definecolor{WHITE}{gray}{1}
 \definecolor{RED}{rgb}{1,0,0}
 \definecolor{GREEN}{rgb}{0,1,0}
 \definecolor{BLUE}{rgb}{0,0,1}
 \definecolor{CYAN}{cmyk}{1,0,0,0}
 \definecolor{MAGENTA}{cmyk}{0,1,0,0}
 \definecolor{YELLOW}{cmyk}{0,0,1,0}
 }

\makeatother

\begin{document}

\title{Correcting detection error in quantum computation and state
engineering \\
through data processing}
\author{C. Shen$^{1,2}$ and L.-M. Duan$^{1,2}$}
\affiliation{$^{1}$Department of Physics and MCTP, University of Michigan, Ann Arbor,
Michigan 48109, USA}
\affiliation{$^{2}$Center for Quantum Information, IIIS, Tsinghua University, Beijing,
China}

\begin{abstract}
Quantum error correction in general is experimentally challenging as it
requires significant expansion of the size of quantum circuits and accurate
performance of quantum gates to fulfill the error threshold requirement.
Here we propose a method much simpler for experimental implementation to
correct arbitrary detection errors. The method is based on processing of
data from repetitive experiments and can correct detection error of any
magnitude, as long as the error magnitude is calibrated. The method is
illustrated with its application to detection of multipartite entanglement
from quantum state engineering.
\end{abstract}

\maketitle

A celebrated achievement in quantum information science is establishment of
the error threshold theorem, which assures that any experimental error below
a certain threshold value can be corrected in quantum computation or
communication \cite{1,1'}. This theorem is important as it indicates there
is no fundamental obstacle to realization of quantum computation with
imperfect experimental devices, as long as the imperfection is small. The
experimental realization of fault-tolerant quantum error correction,
unfortunately, is still challenging. When all the experiment devices are
subject to errors as it is the case for real applications, realization of
fault-tolerance requires significant expansion of the size of quantum
circuits for complicated encoding and accurate performance of quantum gates
to fulfill the error threshold requirement, which is still beyond the
capability of current experimental technology.

In this paper, we propose an alternative method to correct a special but
practically important type of error --- the detection error --- in quantum
information processing. The detection error is a common source of noise in
many quantum information experiments. In particular, it is a significant
obstacle to observation of multipartite entanglement in quantum state
engineering \cite{2}. We show here that this type of error can be corrected
at any magnitude as long as the error magnitude has been calibrated (for
instance, through prior test experiments). The detection error distorts the
experimental data by a transformation that depends on the magnitudes of
various error possibilities. When the relevant error magnitudes have been
calibrated for the detectors by the prior test experiments, the form of the
distortion transformation induced by the detection error is known, and then
we can find a way to inverse this transformation to reconstruct the original
signal. In this way, we can use imperfect detectors to simulate perfect
detectors as long as their imperfection has been calibrated. The proposed
method is straightforward for experimental implementation as it is based on
data processing and removes the difficulty associated with fault-tolerant
quantum encoding. To correct the detection error, we only require to repeat
the same experiments by some additional rounds to have small statistical
error for the inverse transformation. To illustrate its applications, the
method is used to significantly improve the detection of multi-qubit
entanglement and spin squeezing. In many cases, the signal of multipartite
entanglement only becomes visible after the proposed correction of the
detection error, in particular when the number of qubits is large.

Any measurements in quantum information can be reduced to population
measurements in certain bases (including possibly several complementary
bases). If we want to measure properties associated with a state $\rho $
(generally mixed) of $n$ qubits, in each chosen measurement basis, there are
$2^{n}$ possible measurement outcomes. By measurements we determine the
probability $f_{i}$ associated with each outcome $i$ ($i=1,2,\cdots ,2^{n}$%
). For instance, if we repeat the same experiment $N$ times and get the $i$%
th outcome $N_{i}$\ times, we estimate the probability $f_{i}$ by $%
f_{i}=N_{i}/N$ and its standard deviation (the error bar) by $\Delta f_{i}=%
\sqrt{f_{i}(1-f_{i})/N}$ using the binomial distribution. If the detectors
are perfect, the measured probabilities $f_{i}$ just give the distribution $%
g_{i}\equiv \left\langle i\right\vert \rho \left\vert i\right\rangle $ of
the state $\rho $ in the measurement basis $\left\{ \left\vert
i\right\rangle \right\} $. In reality, however, the detectors always have
errors, which distort the distribution $g_{i}$, making the measured
distribution $f_{i}$ significantly different from $g_{i}$. The purpose of
this paper is to show how to reconstruct the real distribution $g_{i}$ from
the measured distorted probabilities $f_{i}$.

We first consider the case where the measurements haven individual
addressing, and each qubit is measured by an independent detector. For
detection on a qubit, the most general error model is characterized by a $%
2\times 2$ matrix
\begin{equation}
D=%
\begin{bmatrix}
1-p_{0} & p_{1} \\
p_{0} & 1-p_{1}%
\end{bmatrix}%
,  \label{1}
\end{equation}%
where $p_{0}$ ($p_{1}$) denotes respectively the error probability that the
detector gives outcome $1$ ($0$) for the input signal of $0$ ($1$). For
simplicity of notation, we assume the error matrix $D$ has the same form for
detection of each qubit (it is straightforward to generalize the formalism
to the case where the error rates $p_{0}$ and $p_{1}$ in the $D$ matrix are
qubit-dependent). Furthermore, we assume $p_{0}$ and $p_{1}$ have been well
calibrated by a prior test experiment. For instance, we may input a known
state to the detector and  can calibrate $p_{0}$ and $p_{1}$ easily from the
measurement data.

For $n$ qubits, the error model for the detection is then characterized by a
$2^{n}\times 2^{n}$ matrix $M=\left[ M_{ji}\right] $, with the element $%
M_{ji}$ corresponding to the probability of recording the outcome $j$ with
the input signal $i$. Assume the detection error rates on different qubits
are independent to each other and the binary string $i$ has $n_{0}$ zeros
and $n_{1}=n-n_{0}$ ones. If we need $\alpha $ flips from $0$ to $1$ and $%
\beta $ flips from $1$ to $0$ to change the string from $i$ to $j$, the
matrix element $M_{ji}$ is given by
\begin{equation}
M_{ji}=\left( 1-p_{0}\right) ^{n_{0}-\alpha }\left( 1-p_{0}\right)
^{n_{1}-\beta }p_{0}^{\alpha }p_{1}^{\beta }.  \label{2}
\end{equation}%
The measured probabilities $f_{j}$ are connected with the real distribution $%
g_{i}$ through the distortion transformation $f_{j}=%
\sum_{i=1}^{2^{n}}M_{ji}g_{i}.$To reconstruct the real signal $g_{i}$ from
the measured distribution $f_{j}$, in principle we only need to inverse the
matrix $M$. However, as $M$ is a huge $2^{n}\times 2^{n}$ matrix, it is not
clear how to inverse this matrix (it is even a question whether the inverse
exists).

Our key observation is that the matrix $M$, with the elements given by Eq.
(2), has a simple tensor product structure. It is straightforward to show by
mathematical induction that
\begin{equation}
M=\bigotimes_{k=1}^{n}D_{k},  \label{4}
\end{equation}%
where all the $D_{k}$ are identical and given by $D$ in Eq. (1).  Therefore,
the inverse can be easily done in an analytic form with
\begin{equation}
M^{-1}=\bigotimes_{k=1}^{n}D_{k}^{-1}=\bigotimes_{k=1}^{n}%
\begin{bmatrix}
1-p_{0}^{\prime } & p_{1}^{\prime } \\
p_{0}^{\prime } & 1-p_{1}^{\prime }%
\end{bmatrix}%
_{i},  \label{5}
\end{equation}%
where the parameters $p_{0}^{\prime }$ and $p_{1}^{\prime }$ are given by
\begin{eqnarray}
p_{0}^{\prime } &=&p_{0}/(p_{0}+p_{1}-1),  \notag \\
p_{1}^{\prime } &=&p_{1}/(p_{0}+p_{1}-1).  \label{6}
\end{eqnarray}%
Note that with the substitution in Eq. (5), $M^{-1}$ and $M$ have the same
form except that $p_{0}^{\prime }$ and $p_{1}^{\prime }$ can not be
interpreted as error rates any more since in general they are not in the
range $\left[ 0,1\right] $. The formula also shows that the inverse
transformation $M^{-1}$ always exists except for the special case with $%
p_{0}+p_{1}=1$.

In some experimental systems we do not have the ability to resolve
individual qubits. Instead, we perform collective measurements on $n$ qubits
by detecting how many qubits (denoted by $j$, $j=0,1,\cdots,n$) are in the
state $\left\vert 1\right\rangle $ in a chosen detection basis (this is
equivalent to measurement of the collective spin operator along a certain
direction). In this case, the detection only has $n+1$ outcomes for an $n$%
-qubit system. For collective measurements on $n$ qubits, the detection
error matrix is represented by an $\left(n+1\right)\times\left(n+1\right)$
matrix $L=\left[L_{ij}\right]$. The matrix element $L_{ij}$ corresponds to
the probability to register outcome $i$ when $j$ qubits are in the $%
\left\vert 1\right\rangle $ state. If the detection error matrix for an
individual qubit is still given by $D$ in Eq. (1), we can directly calculate
$L_{ij}$ from $D$: from signal $j$ to $i$, if $n_{10}$ qubits flip from $0$
to $1$ and $n_{01}$ qubits flip from $1$ to $0$, with the constraints $0\leq
n_{01}\leq j,$ $0\leq n_{10}\leq n-j$ and $n_{01}-n_{10}=j-i$, $L_{ij}$ is
given by

\begin{eqnarray}
L_{ij} &=&\sum_{n_{01,}n_{10}}B(j,p_{1},n_{01})B(n-j,p_{0},n_{10})  \notag \\
&=&\sum_{q}B(j,1-p_{1},q)B(n-j,p_{0},i-q),  \label{7}
\end{eqnarray}%
where $B(n,p,k)\equiv \binom{n}{k}p^{k}(1-p)^{n-k}$ and we have let $%
q=i-n_{10}$, and hence $q$ satisfies the constraint $\max \{0,i+j-n\}\leq
q\leq min\{i,j\}$. As the dimension of the $L$ matrix depends linearly on
the qubit number $n$, it is typically not difficult to numerically calculate
its inverse matrix $L^{-1}$ if $n$ is not very large. As will be shown in
the appendix, there is also a simple analytic formula for $L^{-1}=\left[
L_{ij}^{-1}\right] $: if we denote the dependence of $L_{ij}$ in Eq. (6) on $%
p_{0},p_{1}$ as $L_{ij}=L_{ij}\left( p_{0},p_{1}\right) $, we have
\begin{equation}
L_{ij}^{-1}=L_{ij}\left( p_{0}^{\prime },p_{1}^{\prime }\right)   \label{8}
\end{equation}%
where $p_{0}^{\prime },p_{1}^{\prime }$ are given by the simple substitution
in Eq. (5). With the inverse matrix $L^{-1}$, the real signal $g_{i}$ can be
similarly reconstructed from the measured data $f_{j}$ as $%
g_{i}=\sum_{j}L_{ij}^{-1}f_{j}$.

The above formulation can be extended straightforwardly to qudit ($d$%
-dimensional) systems where the individual detection error matrix $D$ in Eq.
(1) is replaced by a $d\times d$ matrix. For independent detection of $n$%
-qudits, the overall error matrix $M$ still has the tensor-product structure
as shown by Eq. (3), which allows easy calculation of $M^{-1}$ from $D^{-1}$.

With the inverse error matrix $M^{-1}$, it is straightforward to reconstruct
the true distribution $g_{i}$ from the measured data $f_{i}$. The price we
need to pay is that compared with $\Delta f_{i}=\sqrt{f_{i}(1-f_{i})/N}$,
there is an increase of the standard deviation (error bar) $\Delta g_{i}$ in
our estimate of $g_{i}$ by the formula $g_{i}=%
\sum_{j=1}^{2^{n}}M_{ij}^{-1}f_{j}$. With some tedious but straightforward
calculation, we find
\begin{equation}
\Delta g_{i}=\sqrt{[\sum_{j}(M_{ij}^{-1})^{2}f_{j}-g_{i}^{2}]/N}  \label{9}
\end{equation}%
As $M^{-1}=\bigotimes_{k=1}^{n}D_{k}^{-1}$ and $D_{k}^{-1}$ has matrix
element $1-p_{0}^{\prime }\approx e^{p}>1$ (when $p_{0}\sim p_{1}\sim p\ll 1)
$, $M^{-1}$ has matrix element $\sim e^{np}$ which leads to exponential
increase of the error bar $\Delta g_{i}$ with the qubit number $n$. To
maintain the same error bar $\Delta g_{i}$, the number of repetitions $N$ of
the experiment eventually needs to increase exponentially with $n$. For
practical applications, this exponential increase of $N$ by the factor $%
e^{np}$is typically not a problem for two reasons. First, as the detection
error rate $p$ is usually at a few percent level, the exponential factor $%
e^{np}$ remains moderate even for hundreds of qubits. Second, this
exponential increase only applies when we need to measure each element of
the distribution $g_{i}$. In most of quantum information applications, we
only need to measure certain operators which are expressed as tensor
products of a constant number of Pauli operators for different qubits. In
this case, $N$ does not have the exponential increase as we show now.

Suppose we need to measure an operator $\hat{O}$, which is expressed as $%
\hat{O}=\otimes _{k=1}^{n}\sigma _{k}^{\mu _{k}}$, where $\sigma _{k}^{\mu
_{k}}$ is a component of the Pauli matrices when $\mu _{k}=1,2,3$ or the
identity operator when $\mu _{k}=0$. The number of the Pauli matrices $n_{p}$
in the tensor product expansion of $\hat{O}$ is called the support of $\hat{O%
}$. To measure the operator $\hat{O}$, we choose the measurement basis to be
the eigenbasis of $\sigma _{k}^{\mu _{k}}$ for the $k$th qubit. In this
measurement basis, $\hat{O}$ is diagonal with the matrix element $\hat{O}%
=\otimes _{k=1}^{n}$diag$\left( \sigma _{k}^{\mu _{k}}\right) $, where diag$%
\left( \sigma _{k}^{\mu _{k}}\right) =\left[ 1,1\right] $ for $\mu _{k}=0$
and diag$\left( \sigma _{k}^{\mu _{k}}\right) =\left[ 1,-1\right] $ for $\mu
_{k}=1,2,3$. Under the distribution $g_{i}$, the expectation value of $\hat{O%
}$ is given by $\left\langle \hat{O}\right\rangle =\sum_{i}\hat{O}%
_{i}g_{i}=\sum_{i}\hat{O}_{i}\sum_{j}M_{ij}^{-1}f_{j}=\sum_{j}(\sum_{i}\hat{O%
}_{i}M_{ij}^{-1})f_{j}\equiv \sum_{j}\hat{O_{j}^{c}}f_{j}$, where $\hat{O}%
_{i}$ denotes the diagonal matrix element of $\hat{O}$. Therefore, by
defining a corrected operator $\hat{O}^{c}$, we can get the true expectation
value $\left\langle \hat{O}\right\rangle $ directly from the experimental
data $f_{j}$. Using the relation $M^{-1}=\bigotimes_{k=1}^{n}D_{k}^{-1}$, $%
\hat{O}^{c}$ is expressed as $\hat{O^{c}}=\bigotimes_{k=1}^{n}\left[ \text{%
diag}(\sigma _{k}^{\mu _{k}})D_{k}^{-1}\right] .$ For $\mu _{k}=1,2,3$,

\begin{align}
\text{diag}(\sigma _{k}^{\mu _{k}})D_{k}^{-1}& =\left[
\begin{array}{cc}
1 & -1%
\end{array}%
\right] \left[
\begin{array}{cc}
1-p_{0}^{\prime } & p_{1}^{\prime } \\
p_{0}^{\prime } & 1-p_{1}^{\prime }%
\end{array}%
\right]   \notag \\
=& \left[
\begin{array}{cc}
(1-2p_{0}^{\prime }) & -(1-2p_{1}^{\prime })%
\end{array}%
\right]   \label{10}
\end{align}%
and for $\mu _{k}=0$, diag$(\sigma _{k}^{\mu _{k}})D_{k}^{-1}=[1,1]$. For
simplicity of notation, we take $p_{0}=p_{1}=p$. In this case, diag$(\sigma
_{k}^{\mu _{k}})D_{k}^{-1}=\left( 1-2p\right) ^{-1}$diag$(\sigma _{k}^{\mu
_{k}})$ for $\mu _{k}=1,2,3$, and the corrected operator $\hat{O}^{c}$ is
related with the original operator $\hat{O}$ by a simple scaling
transformation $\hat{O}^{c}=\left( 1-2p\right) ^{-n_{p}}\hat{O}$. \ The
scaling transformation is independent of the qubit number $n$, so the error
bar of $\left\langle \hat{O}\right\rangle $ does not have exponential
increase with $n$ when the operator $\hat{O}$ has a constant support $n_{p}$.

The scaling transformation also applies to collective operators, but some
caution needs to be taken for calculation of their variance. For instance,
if we take the collective spin operator $J_{z}\equiv \sum_{k=1}^{n}\sigma
_{k}^{z}/2$, it is easy to see that $J_{z}^{c}=\left( 1-2p\right) ^{-1}J_{z}$%
\ as each of the terms of $J_{z}$ has support $n_{p}=1$. However, as $%
J_{z}^{2}\equiv n/4+\sum_{k\neq l}\sigma _{k}^{z}\sigma _{l}^{z}/4$ which
has non-uniform support for its superposition terms, one finds that $\left(
J_{z}^{2}\right) ^{c}=n/4+\left( 1-2p\right) ^{-2}\left(
J_{z}^{2}-n/4\right) =\left( 1-2p\right) ^{-2}\left[ J_{z}^{2}-np\left(
1-p\right) \right] $. With this transformation, we can correct the
distortion to the spin squeezing parameter by the detection error. Assume
that the mean value of $\left\langle \mathbf{J}\right\rangle $ is along the $%
x$-direction with $\left\langle \mathbf{J}\right\rangle =\left\langle
J_{x}\right\rangle $ and the squeezing is along the $z$-direction. The
squeezing parameter is given by $\xi =\sqrt{n\left\langle
J_{z}^{2}\right\rangle /\left\langle J_{x}\right\rangle ^{2}}$ \cite{3}.
Using the transformation for $\left( J_{z}^{2}\right) ^{c}$ and $J_{x}^{c}$,
we find that
\begin{equation}
\xi ^{c}=\sqrt{n\left\langle \left( J_{z}^{2}\right) ^{c}\right\rangle
/\left\langle J_{x}^{c}\right\rangle ^{2}}=\sqrt{\xi ^{2}-\xi _{d}^{2}}
\label{11}
\end{equation}%
where $\xi _{d}^{2}=n^{2}p(1-p)(1-2p)^{-2}\left\langle J_{x}\right\rangle
^{-2}$ is the contribution to $\xi ^{2}$ by the detection noise. After
correction of the detection error, $\xi ^{c}$ gets significantly smaller
compared with $\xi $ in particular when the qubit number $n$ is large, and
thus can be used to verify a much bigger entanglement depth using the
criterion in Ref. \cite{4}. From Eq. (10), we find that the variation $%
\Delta \xi ^{c}/\Delta \xi =\xi /\xi ^{c}$. As typically $\xi \gg \xi ^{c}$,
the error bar for $\xi ^{c}$ after correction of the detection error gets
significantly larger, and we need to correspondingly increase the rounds of
the experiment $N$ to reduce the statistical error.

To illustrate application of the error correction method here, as an
example, we apply it to detection of genuine multi-partite entanglement in
graph states. For a graph state $\left\vert G_{n}\right\rangle $ of $n$
qubits associated with a $q$-colorable graph $G$, the genuine $n$-party
entanglement can be detected with the following witness operator \cite{5}

\begin{equation}
W_{G_{n}}=3\mathbb{I}-2\left[ \sum_{l=1}^{q}\left( \prod_{k\in Q_{l}}\left(
S_{k}+\mathbb{I}\right) /2\right) \right]   \label{12}
\end{equation}%
where $Q_{l}$ denotes the set of qubits with the $l$th color $\left(
l=1,2,\cdots ,q\right) $, $\mathbb{I}$ is the identity operator, and $S_{k}$
is the stabilizer operator for the $k$th qubit (which is a tensor product of
the Pauli operators \ $\sigma _{k}^{x}$ for the $k$th qubit and $\sigma
_{k^{\prime }}^{z}$ for all it neighbors $k^{\prime }$ in the graph $G$). A
state $\rho $ has genuine $n$-qubit entanglement if $tr\left( \rho
W_{G_{n}}\right) =\left\langle W_{G_{n}}\right\rangle <0$. For an ideal
graph state, all its stabilizer operators $S_{k}$ have expect values $%
\left\langle S_{k}\right\rangle =1$. With detection error, the value of $%
\left\langle S_{k}\right\rangle $ gets significantly degraded. As an
example, Fig. 1 shows the values of all $\left\langle S_{k}\right\rangle $
for two particular $2$-colorable graph states: a $10$-qubit GHZ\ state (GHZ$%
_{10}$) and a linear cluster state (LC$_{10}$). We assume $3\%$ detection
error with $p_{0}=p_{1}=p=0.03$ for each qubit. With a known magnitude $p$,
the detection error can be easily corrected by a scaling transformation $%
S_{k}^{c}=\left( 1-2p\right) ^{-n_{pk}}S_{k}$, where $n_{pk}$ is the support
of the corresponding stabilizer operator $S_{k}$. Fig. 1 shows that after
error correction, $\left\langle S_{k}^{c}\right\rangle $ is almost unity.
Its error bar increases a bit after the correction, but is still small. To
show the influence on the entanglement detection, we assume the
experimentally prepared graph state $\rho _{ex}$ corresponds to the ideal
target state $\rho _{id}$ distorted by small depolarization noise
independently acting on each qubit, so $\rho _{ex}=\widehat{\$}\left( \rho
_{id}\right) $, where the noise super-operator $\widehat{\$}%
=\bigotimes_{k=1}^{n}\widehat{\$}_{k}$ and $\widehat{\$}_{k}\left( \rho
_{id}\right) =(1-3p_{n}/4)\rho _{id}+p_{n}/4\sum_{\mu =x,y,z}\sigma
_{k}^{(\mu )}\rho _{id}\sigma _{k}^{(\mu )}$ \cite{6}. In Fig. 2, we show
the witness $\left\langle W_{G_{n}}\right\rangle $ as a function of the
preparation error rate $p_{n}$, both before and after correction of the
detection error (with an error rate $p=3\%$). For both GHZ$_{10}$ and LC$%
_{10}$ states, without correction of the detection error, we cannot detect
any $n$-qubit entanglement even for a perfectly prepared state with $p_{n}=0$%
. After correction of the detection error, we can confirm genuine $n$-qubit
entanglement as long as the preparation error $p_{n}\lesssim 5\%$. So,
correction of the detection error significantly improves the experimental
performance, and the improvement gets more dramatic when the qubit number
increases.

Finally, we briefly comment on the sensitivity of our error correction
method to calibration of the detection error. In this method, the error
magnitude $p$ (or magnitudes $p_{i}$, $i=0,1,\cdots ,$ for general cases) is
assumed to be known. If we have a relative error $e$ in calibration of the
magnitude $p$, .i.e., $\delta p/p\sim e$, the scaling transformation on the
detected operator $\hat{O}$ leads to an relative error in the observed
quantity $\delta \left\langle \hat{O}\right\rangle /\left\langle \hat{O}%
\right\rangle \sim 2n_{p}\delta p\left( 1-2p\right) ^{-1}\sim 2n_{p}pe$. As
long as $2n_{p}p\precsim 1$, which is typically the case as $p\ll 1$, the
relative error actually gets reduced and the method here can tolerate some
uncertainty in calibration of the error magnitude $p$.

\begin{figure}[tbp]
\label{Fig1} \includegraphics[height=4cm,width=9cm]{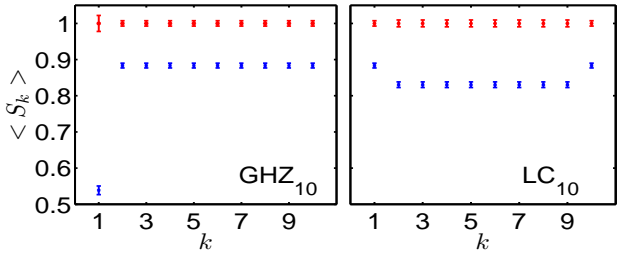}
\caption{Values of stabilizers before (lower points) and after (upper
points) correction of the detection error (with the error rate $p=0.03$) for
the $10$-qubit GHZ state (GHZ$_{10}$) and the linear cluster state (LC$_{10}$%
). Error bars account for the statistical error by assuming $N=5000$
independent measurements in each detection setting.}
\label{fig:stabilizers}
\end{figure}

\begin{figure}[tbp]
\label{Fig2} \includegraphics[height=4cm,width=9cm]{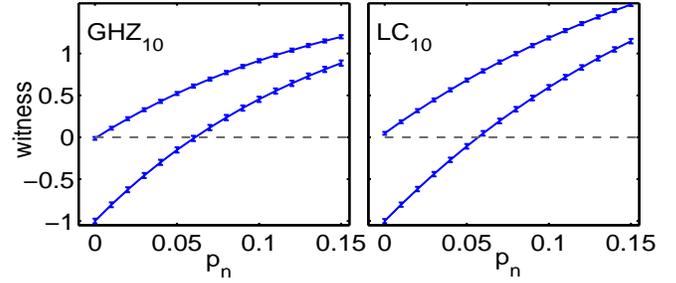}
\caption{The entanglement witness $\left\langle W_{G_{n}}\right\rangle $
under different state preparation errors $p_{n}$ for GHZ (GHZ$_{10}$) and
cluster (LC$_{10}$) states before (upper lines) and after (lower lines)
corection of the detection error (error rate $p=0.03$). The error bars are
obtained by assuming $N=5000$ rounds of measurements in each detection
setting. }
\label{fig:stab_witness}
\end{figure}

In summary, we have shown a method to correct any detection error through
simple processing of the experimental data. The method applies to
measurements in general many-particle settings, with or without separate
addressing. Different from the conventional quantum error correction, this
method does not require encoding or change of the quantum circuits and works
under arbitrary magnitudes of the detection noise, as long as the error
magnitude has been calibrated. The cost of this method is moderate as it
only requires repetition of the same experiment by some additional rounds to
gain enough statistics and thus the method can readily apply to many
experimental settings.

\textbf{Appendix:Proof of Eq (7).} We can relate the $L$ matrix to the $M$
matrix defined in Eq. (3). Denote the space of $n$-bit binary strings with $i
$ bits of $1$ as $S_{i}$, and $S_{i}$ has dimension $\binom{n}{i}$. The
matrix element $M_{\sigma \rho }$ represents the probability of recording a $%
n$-bit binary string $\rho $ as $\sigma $, and $L_{ij}$ is the probability
of recording a signal $\rho \in S_{j}$ as any string in the $S_{i}$ space.
As a collective measurement does not distinguish the binary strings in the
same space $S_{i}$, $L_{ij}$ is related to $M_{\sigma \rho }$ by $%
L_{ij}=\sum_{\sigma \in S_{i}}M_{\sigma \rho }.$The probability $L_{ij}$ is
apparently independent of the exact form of $\rho $, as long as $\rho $
belongs to the space $S_{j}$, so we can pick up any $\rho \in S_{j}$ in $%
L_{ij}=\sum_{\sigma \in S_{i}}M_{\sigma \rho }$ without alternation to the
result of summation. From Eq. (4), we know $M_{\mu \nu }^{-1}=M_{\mu \nu
}\left( p_{0}^{\prime },p_{1}^{\prime }\right) $. Let us define
\begin{equation*}
N_{jk}\equiv \sum_{\mu \in S_{j}}M_{\mu \nu \text{ }}^{-1}=\sum_{\mu \in
S_{j}}M_{\mu \nu \text{ }}\left( p_{0}^{\prime },p_{1}^{\prime }\right)
=L_{jk}\left( p_{0}^{\prime },p_{1}^{\prime }\right) ,
\end{equation*}%
where $\nu $ is an arbitrary element in $S_{k}$. Now we show that $N$ gives
inverse of the matrix $L$: \
\begin{eqnarray*}
&&\sum_{j}L_{ij}N_{jk}=\sum_{j}\sum_{\sigma \in S_{i},\mu \in
S_{j}}M_{\sigma \rho }M_{\mu \nu }^{-1} \\
&=&\sum_{\sigma \in S_{i}}\sum_{j}\sum_{\mu \in S_{j}}M_{\sigma \mu }M_{\mu
\nu }^{-1}=\sum_{\sigma \in S_{i}}\delta _{\sigma \nu }=\delta _{ik}
\end{eqnarray*}%
In the second line, we have changed the subscript $\rho $ in $M_{\sigma \rho
}$ to $\mu $ as both $\rho ,\mu $ belong to $S_{j}$. This proves Eq. (7)\ in
the text.

This work was supported by the NBRPC(973 Program) 2011CBA00300
(2011CBA00302), the IARPA MUSIQC program, the DARPA OLE program, the ARO and
the AFOSR MURI program.

\end{document}